# Full-range optical imaging of planar collagen fiber orientation using polarized light microscopy

Michaela Turčanová[1,*], Martin Hrtoň[2], Petr Dvořák[2], Kamil Novák[1], Markéta Hermanová[3], Zdeněk Bednařík[3], Stanislav Polzer[4], Jiří Burša[1]

[1] Brno University of Technology, Faculty of Mechanical engineering, Institute of Solid Mechanics, Mechatronics and Biomechanics, Technická 2896/2, Brno, 616 69, Czech Republic

[2] Brno University of Technology, Faculty of Mechanical engineering, Institute of Physics engineering, Technická 2896/2, Brno, 616 69, Czech Republic

[3] 1st Department of Pathology, St. Anne's University Hospital Brno and Faculty of Medicine, Masaryk University, Pekařská 664/53, 656 91, Brno, Czech Republic

[4] Technical University Ostrava, Faculty of mechanical engineering, Department of Applied Mechanics, 17. listopadu 15, Ostrava, 708 33, Czech Republic

[*]Corresponding author: turcanova@fme.vutbr.cz

**Abstract:** A novel method for automated assessment of directions of collagen fibers in soft tissues is presented. It is based on multiple rotated images obtained via polarized light microscope with perpendicular and inclined polarizers and thus it breaks the limitation of 90° periodicity of polarized light intensity and evaluates the fiber orientation over the whole 180° range accurately and quickly. After having verified the method, we used histological specimens of porcine Achilles tendon and aorta to validate the proposed algorithm and to lower the number of rotated images needed for evaluation. Our algorithm is capable to analyze $5 \cdot 10^5$ pixels in one micrograph in a few seconds and is thus a powerful and cheap tool for histological image analysis of such samples promising a broad application.

**Keywords**: soft tissue, histological slice, collagen, fiber direction, polarized light microscopy, phase correlation, histogram.

## 1. Introduction

Collagen fibers and their structure are responsible for the distinctive mechanical response of many soft biological tissues such as skin, tendons, cartilages, myocardium or arteries [1]. Similarly, the observed differences in mechanical behavior (flexibility and strength) between healthy and pathological arteries (e.g. abdominal aortic aneurysm) are largely related to changes in collagen organization [2–4]. Consequently, knowledge of collagen structure in soft tissues is crucial for a deeper understanding of their mechanical behavior and for their structure-based constitutive description [5] which is decisive for clinical applications of computational models. For instance, computational modelling offers the best assessment of the rupture risk of aortic aneurysm or atherosclerotic plaque [4,6–8] and tends to be exploited in timing of surgeries.

There are several approaches how to visualize collagen fibers in soft tissues but each has some limitations. For example, the small-angle X-ray scattering (SAXS) [9] suffers from disadvantages such as degradation of organic fibers, complexity of optical elements, and need for safety precautions to prevent inadvertent irradiation. Another frequently used method, small-angle light scattering (SALS) [10], is precise but also highly time-consuming. Fluorescence microscopy-based methods, such as confocal laser scanning microscopy (CLSM) [11,12] or multiphoton microscopy (MM) [13], are precise and fast but require expensive microscopic equipment and mostly a complex sample staining with combination of adhesion protein and a special fluorescent dye. Non-centrosymmetric crystalline triple helix structure of collagen allows to emit second harmonic generation (SHG) light used in multiphoton or confocal microscopy [14]. In [15] polarization-resolved SHG microscopy was used to show the orientation of collagen in human facial skin in vivo. This method has proven to be particularly suitable for the observation and analysis of the uniaxial orientation of collagen, e.g. in the skin or tendon but is not very suitable for more complex tissues (vessel wall) [15]. Here quantitative polarized light microscopy (QPLM) [16–18] can be more effective, it requires, however, a sophisticated and rather expensive equipment: a modified confocal microscope with quarter-wave plate and rotating analyzer. The above studies demonstrated the ability of QPLM to determine the orientation of collagen, but they did not address the dispersion of fibers and its subsequent use in constitutive models. For this purpose, fast Fourier transform is often applied [19–21] but it was shown that although it evaluates the dominant fiber directions correctly [19], it requires calibration by ground truth data to correctly capture the dispersion of fiber directions [20] which represents a persisting time consuming disadvantage of this approach.

In contrast, classical polarized light microscopy (PLM) is a simple and widely used optical method preferred in analyses of collagen structure because it allows evaluation over a large sample area and does not require any expensive equipment or complex sample preparation [4,19,22–24]. It is based on two mutually perpendicular (crossed) polarizers that transmit only a portion of the polarized light scattered by a birefringent specimen. Nevertheless, this method has two limitations: firstly, in case of manual evaluation, it is very slow. Consequently, a typical number of manually evaluated points (volume elements) per image is of the order of tens [4,19,24] which is not sufficient to extract rigorous information especially regarding dispersion of collagen fibers. Secondly, two mutually perpendicular structures cannot be distinguished from each other due to its intrinsic 90° uncertainty originating from the periodic nature of the pixel angular intensity profile (i.e. pixel intensity as a function of the sample rotation).

In this paper we present a new method for automatic evaluation of the orientation and dispersion of collagen fibers using PLM. We have overcome the angular limitation given by the periodicity of light intensity at crossed filters and proposed an automatic algorithm that is capable to evaluate the orientation in each pixel of the micrograph.

## 2. Materials and Methods

### 2.1 Preparation of samples

The proposed method was validated using porcine Achilles tendon and porcine aortic wall. Specimens were harvested in a local slaughterhouse from 10 months old pigs weighting 105-120 kg. The preparation of histological sections was done in St. Anne's University Hospital in Brno.

Achilles tendon consists of unidirectionally oriented collagen fibers being prone to undulation if an unloaded tendon is used to harvest the specimen. Cuboid specimens of ca 20×8×8 mm were cut off and fixed in 10 % formaldehyde solution at room temperature for 24 hours. After that the samples were dehydrated and embedded in paraffin. Microtome was used to cut 5 μm thick slices parallel to the preferred (longitudinal) direction of collagen fibers. Finally, every slice was stained with a special dye Picro Sirius Red (PSR) (0.1 %) to intensify the birefringence of collagen [25].

Aortic specimens with dimensions 18×18 mm were taken from the anterior wall of straight part of thoracic aorta. They were flattened with four pins on a plywood and the slices were cut in circumferential-axial plane.

*2.2 Experimental setup*

All the histological slices were scanned with an upright microscope (Padim, Drexx s.r.o., CZ) in a transmission configuration equipped with a digital camera (Bresser microcam 5 megapixel, Bresser GmbH, Germany) and standard 2D rotary stage (Padim, Drexx s.r.o., CZ). All the images were recorded under identical illumination conditions and magnification (10× objective, numerical aperture 0.17; 10× ocular); the trimmed images had the size of 960 × 960 pixels (pixel size 0.73 μm). A halogen lamp with a power of 100 W was used as a white light source.

*2.3 Procedure description and the proposed automated algorithm*

As mentioned above, the standard PLM setting with cross-polarized filters, i.e. with perpendicular orientation of polarizer 1 and polarizer 2 (called analyzer below) is shown in Fig. 1a). Consequently, no light can pass through both polarizers, unless its polarization state is changed by a birefringent (optically anisotropic) material located in the light trajectory between both polarizers. The birefringent material rotates the plane of polarized light towards its major axis of anisotropy so that this plane is no longer perpendicular to the analyzer and some portion of the light can pass through. Zero intensity (fiber extinction) occurs when the fiber is oriented in parallel with either the polarizer or the analyzer. In contrast, the intensity of the passing light is maximal for the fiber angle of 45° with respect to both polarizers. Another important feature of collagen is its diattenuation, a total anisotropic attenuation of light caused by absorption and scattering (see Fig. 1b). Intensity of the transmitted light depends on its polarization and is maximal and minimal for the light polarized in two mutually perpendicular directions [26]. Proposed method exploits combination of these two effects to switch the angular periodicity of the transmitted light from 90° to 180° for non-perpendicular polarizers, as illustrated with images of porcine tendon in Fig. 1c) where angle $\delta$ represents the deflection from the cross-polarized setting.

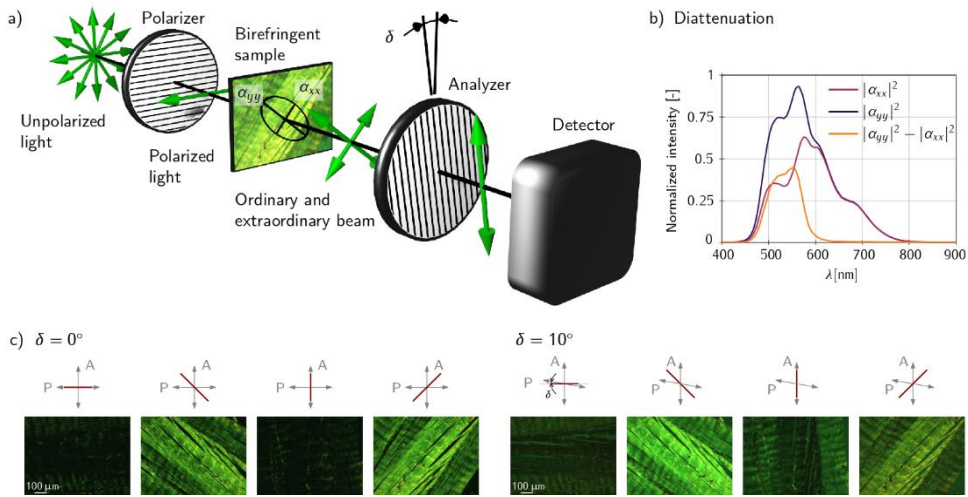

Fig. 1: a) Experimental scheme of a polarized light microscope with unidirectional fibers. b) Spectral dependence of the transmission matrix components $\alpha_{xx}$, $\alpha_{yy}$ and their difference. Anisotropy of collagen fibers is most pronounced at wavelengths between 550 and 570 μm, as illustrated by the difference between $|\alpha_{xx}|^2$ and $|\alpha_{yy}|^2$ representing light intensities of ordinary and extraordinary beam due to parallel setting of polarizers. c) Images of a sample of porcine Achilles tendon with polarizers being either perpendicular (four images in the left) or deflected by angle $\delta$ (four images in the right).

In contrast to manual measurements, the automated algorithms record the polarized light intensities under different rotations of the specimen and subsequently fit the measured data with a theoretical intensity curve. The proposed algorithm builds on our approach reported in [27] which exploited a well-known phase-correlation and 90° cosine-like periodicity of the polarized light intensity under 2D (in-plane) rotation of the specimen. Within this approach, three rotated images were needed to fit the position of the purely cosine-like normalized intensity curve which, however, cannot suffice here due to non-sinusoidal character of the periodic function.

In the first step we recorded a set of 18 images rotated per 10° with cross-polarized setting (i.e. $\delta = 0°$, see Fig. 2a). The algorithm uses phase correlation procedure to rotate and shift all the images to aligned positions and then it evaluates their light intensities in corresponding pixels of all the rotated images. These values are subsequently fitted with the following cosine-like function:

$$I(\theta) = \frac{1}{2} - \frac{1}{2}\cos[4(\theta - p)], \quad (1)$$

where $\theta$ denotes the rotation angle of the sample with respect to the polarizer and $p$ represents the angular (phase) shift of this function determining the fiber orientation in the investigated pixel. This parameter is fitted to obtain the best correlation with the measured values which is in turn achieved by maximizing Pearson's correlation coefficient (see [27] for more details).

The minima in the cosine-like function represent the angles, at which the fiber in the investigated pixel is oriented along the polarization axis of either the polarizer or the analyzer. As one cannot distinguish between these two perpendicular orientations, this ambiguity is resolved by setting $\neq 0°$, i.e. by employing inclined polarizers. A full mathematical derivation of the intensity profile in this setting can be found in Supplemental document, here we present only the final approximate formula for small values of $\delta$:

$$I(\theta) \approx \frac{1}{8}|\alpha_{xx} - \alpha_{yy}|^2[1 - \cos(4(\theta - p))] \\ - \frac{1}{2}\delta\left(|\alpha_{xx}|^2 - |\alpha_{yy}|^2\right)\sin(2(\theta - p)) \quad (2)$$

where $\alpha_{ii}$ represents attenuation in direction $i$ (see Fig. 1b).

The first term describes the original intensity profile with 90° periodicity, while the second term possesses periodicity twice as large and overcomes thus the limitations inherent to the cross-polarized configuration. Note that apart from the mutual inclination of the polarizers, the sample is also required to exhibit significant diattenuation ($|\alpha_{xx}|^2 - |\alpha_{yy}|^2 \neq 0$), a prerequisite satisfied in collagen fibers.

Therefore, the second step of our algorithm consists of recording another set of 18 images with inclined polarizers ($\delta = 10°$), which are then used to select the true orientation of fibers in our specimen from the two possible options found in the previous step as described below.

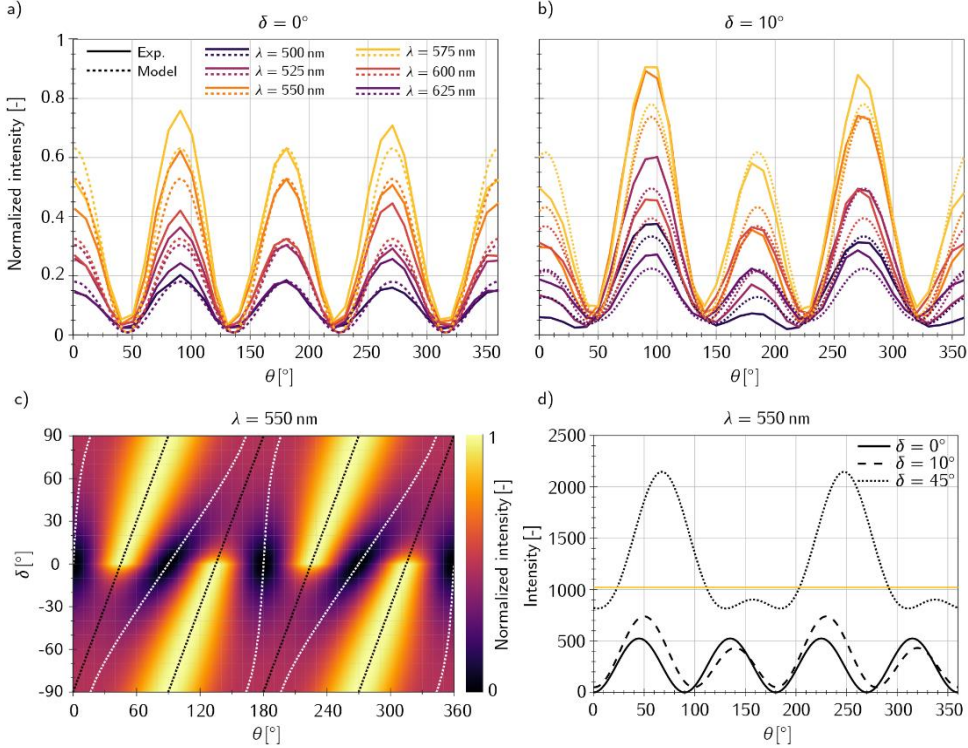

Fig. 2: a),b) Dependences of the intensity of transmitted polarized light on the fiber angle $\theta$ for six different wavelengths $\lambda$ and two different analyzer rotation angles $\delta$; the experimental results agree perfectly with the created model. c) Intensities of green light with wavelength $\lambda = 550$ nm for different analyzer angles $\delta$ and sample rotation angles $\theta$ (the fiber angle $p$ is implicitly set to 0°). Black and white dotted lines represent maxima and minima of the intensity calculated for a certain value of $\delta$, respectively. d) Horizontal sections through the graph c), representing functions of sample rotation angle $\theta$ for three different analyzer angles $\delta$. Solid line ($\delta = 0°$) shows 90° periodicity while the other angles show 180° periodicity with each second maximum being highly suppressed. Extreme differences in intensities between $\delta = 0°$ and $\delta = 45°$ disable practical application of this combination of angles, as the intensity curve for $\delta = 45°$ is above the pixel saturation value for most $\theta$ angles (shown by the yellow line in the graph). However, when using another camera with higher saturation level, another $\delta$ value can be used and introduced easily into the algorithm.

Examples of the intensity curves, both measured (using a fiber optic spectrometer BW-VIS2) and calculated (using the full expression given by eq. (S10) in Supplemental document), are presented in Fig. 2a) and b). For $\delta = 0°$, the intensity profiles possess the predicted 90° periodicity (with slight deviations of experimental curves due to noise), while a pronounced difference occurs between even and odd local maxima for $\delta = 10°$, enabling us to distinguish between the two mutually perpendicular orientations. This difference between the neighboring local maxima depends on the wavelength; on the basis of measurements and simulations, the wavelength of 550 nm (green color) was chosen as the best candidate, since it provides the largest difference between $|\alpha_{xx}|^2$ and $|\alpha_{yy}|^2$, as shown in Fig. 1b). To choose optimal $\delta$ for this wavelength, we calculated the dependence of light intensity on sample rotation angle $\theta$ and analyzer angle $\delta$ (for its full range between -90° and +90°) using the measured transmittance data from Fig. 1b); the resulting intensity map is plotted in Fig. 2c).

This map shows that for $\delta \gtrsim 5°$ each second maximum is significantly reduced and thus the periodicity changes to 180°. Cuts throughout the intensity map in Fig. 2c) for analyzer angles $\delta$ of 0°, 10°, and 45° show this phenomenon, making the minima more flat and hardly detectable (see Fig. 2d). Therefore, the algorithm exploits the larger (non-reduced) maximum instead. As soon as the larger maximum is found, the angle of minimum intensity corresponds to the position of the minimum to the left (right) of this maximum for positive (negative) analyzer angle $\delta$, respectively (see Fig. 2c). We have chosen $\delta = 10°$ because a higher deviation may causes overexposure of the image (see Fig. 2d) where a high increase in light intensity is shown for deflection $\delta = 45°$. Moreover, the shift between maxima of the π/2-periodic and π-periodic intensity functions increases with angle $\delta$ and reaches values close to 90° for $\delta = 45°$ (see Fig. 2c) and d)); this could make the switching between left and right minima sensitive to additional errors. Note that one could, in principle, extract the fiber orientation solely from the set of images with inclined polarizers, i.e. from the dashed or dotted curve in Fig. 2d). However, the inaccuracies in the evaluation of the relatively flat minima in the non-sinusoidal function, together with inaccuracies in the measured diattenuation parameters of the sample render this approach rather unreliable. This apparent susceptibility to errors was the main reason standing behind our decision to adopt the more robust two-step procedure described above.

## 3. Results

### 3.1 Verification of the algorithm

As the functionality of the original algorithm with π/2 periodicity was already validated in [27], it now remains to verify to the proposed extension that should allow us to distinguish between both mutually perpendicular directions. To this end, a sample of porcine Achilles tendon was chosen because the orientation of its unidirectional slightly wavy fibers can be easily measured manually (see Fig. 3a). The light intensity curves were evaluated with inclined polarizers for three selected pixels, where the fiber in pixel number 3 is nearly perpendicular to the fiber orientation dominating the majority of the sample. The resulting histogram (Fig. 3c) is interpolated with the following von Mises distribution function [28]:

$$\rho(\varphi) = \frac{\exp(b \cdot \cos(\varphi - \mu))}{2 \cdot \pi \cdot I_0(b)}, \qquad (3)$$

where $b$ is the concentration parameter, $\varphi$ is the pixel-wise evaluated fiber angle, $\mu$ is the mean direction of the distribution and $I_0(b)$ is Bessel function. Quality of this approximation is evaluated by means of coefficient of determination $R^2$. This approximation and its parameters are shown with the histograms. The main maxima which grow in magnitude as we increase the inclination of the polarizers are recognizable in the waveforms in Fig. 3d). Following the two-step procedure introduced earlier, the orientation of the fiber (measured as the deviation from the horizontal direction) is determined from the π/2-periodic intensity function (recorded with perpendicular polarizers) as the angle of the minimum located just to the left of the main (larger) maximum in the corresponding π-periodic intensity function (recorded with inclined polarizers) recognizable in the waveforms in Fig. 3d). Thus, the second set of images measured with inclined polarizers serves only as a logic gate that allows us to choose between the two mutually perpendicular orientations of the fibers determined in the first step.

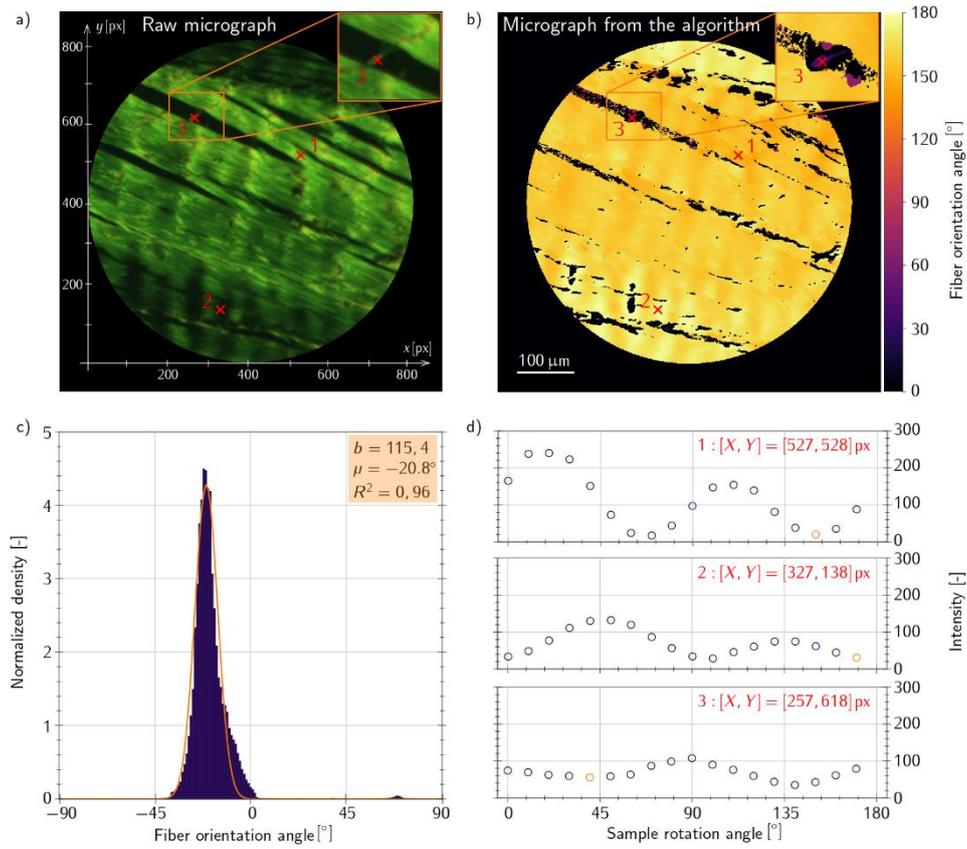

Fig. 3: Verification of the algorithm at three different locations of porcine Achilles tendon sample. Figure a) shows a micrograph (the part with pixel 3 is zoomed in upper right corner) with a setting for green color and $\delta = 0°$. The micrograph with directions evaluated by the proposed algorithm is shown in color scale in b). The histogram obtained for all pixels is interpolated in c) with the von Mises distribution function and its parameters are listed in the legend. Figure d) shows the light intensity dependence on the rotation angle in the selected three pixels of the sample with inclined polarizers. The minimum to the left of the main maximum indicates the angle of the fiber (red circle) which is detected accurately from the $\pi/2$ periodic intensity function from the first set. In this way also the pixel 3 is evaluated correctly and its highly different orientation which is visible in the histogram.

*3.2 Validation of the algorithm*

Fig. 4 shows the validation results, i.e. comparison between the automated algorithm and manual measurements. The results of the manual measurement, with some 300 square shaped evaluated areas (each of approx. 39 × 39 pixels), are presented in histograms. Deviations between the parameters of the manual and automated evaluation (between 4° and 6° for the mean angle) can be attributed to the (by orders) smaller numbers of areas evaluated manually. This is indicated by the lower $R^2_{man}$ values (coefficient of determination of the fit to manual measurements) while for the automated algorithm the approximation of histograms is perfect ($R^2_{alg} > 0.96$). The manual histograms suffer evidently from low numbers of measurements and consequently from an impact of noise, as confirmed by a better accordance of two less accurate fitting curves of manual data with the fit of the algorithm data (represented by $R^2_{man-alg}$ in Fig. 4) than with the manual measurements themselves ($R^2_{man}$).

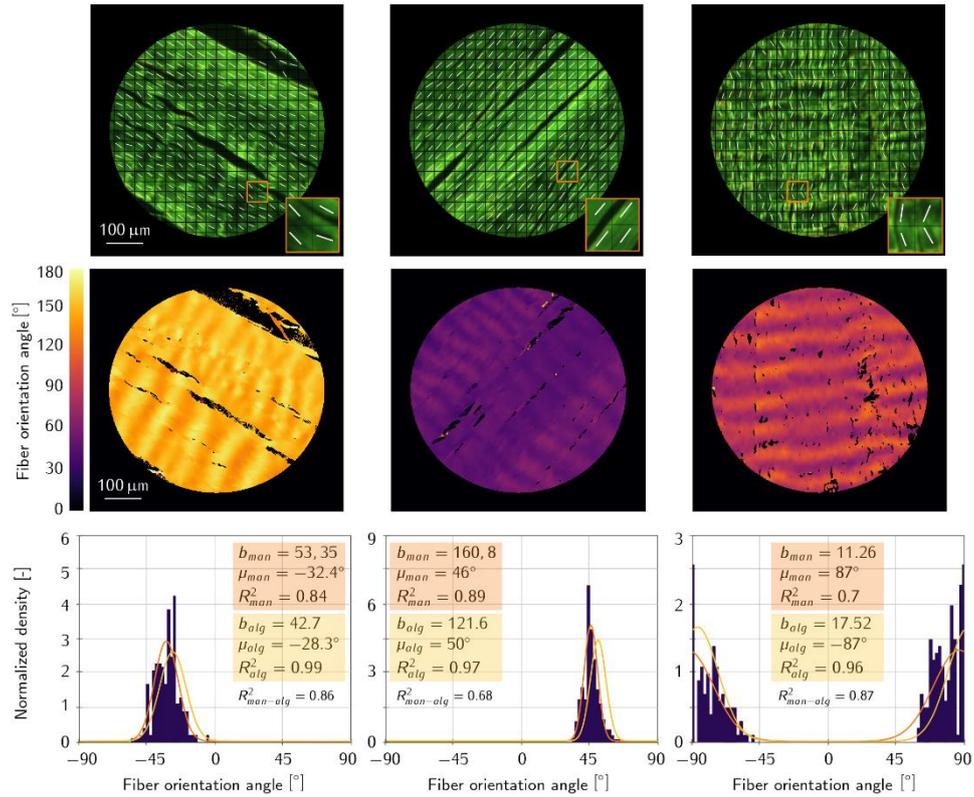

Fig. 4: Validation of the algorithm with porcine Achilles tendon using manual measurement of fiber angles in the marked areas. The micrographs in the upper row show c. 300 marked areas with the fiber angles determined by the operator. The images in the middle row show (in color scale) the fiber orientation evaluated by the algorithm. Histograms from the manual measurements are displayed in the bottom row, completed in the legends with parameters of the von Mises distributions fitted to the manual measurements (with subscripts man) and to the algorithm results (with subscripts alg).

*3.3 Number of figures needed for detection of orientation of collagen fibers*

The presented procedure exploits two sets of 18 images with rotation step of 10°. To reduce the time-consumption of this approach, we tested what is the minimum number of images necessary to detect the correct orientation of collagen fibers without a significant loss of information or accuracy. For this purpose, we compared results obtained with the proposed algorithm for different selections from the measured data set for the porcine Achilles tendon (see Fig. 5). The resulting histograms in the lower left corner of the micrographs are fitted with von Mises distribution function for which mean direction *μ*, concentration parameter *b*, and coefficient of determination $R^2$ are listed in the figures. The percentage indicates the portion of evaluated pixels; as Achilles tendons consist dominantly of collagen fibers, all pixels should be evaluated except for holes between the individual fiber bundles or tissues other than collagen.

Although in all the compared variants the mean angle is evaluated correctly, there are significant differences in the other parameters of the distribution. While the first three micrographs are almost identical in all parameters, the last histogram (based on 3 images only) suffers from a lower percentage of evaluated pixels, a significant number of erroneously evaluated angles (rotated by 90°), and consequently a different concentration parameter *b*. For the histogram based on 4 images, these errors are also significant although lower. In contrast, 5 images seem to give correct parameters of the distribution but a higher percentage of not evaluated pixels (i.e. the pixels which fell out because their Pearson coefficient was below a set minimum threshold and the fiber orientation could not be therefore reliably determined) indicates a risk of errors when evaluating less homogeneous tissues. Consequently, 6 images (corresponding to 30° rotation step) were decided to represent a necessary minimum that need to be recorded with both perpendicular and inclined polarizers. This may seem to contradict our earlier statement that 3 images are sufficient for the correct evaluation of minima within the π/2-periodicity of light intensity profile recorded with perpendicular polarizers [27]. This inconsistency is due to our decision to keep the number of images in both sets (with perpendicular and inclined polarizers) the same in order to avoid operator errors and increase the accuracy. Although we did not explore it, the variant with 3 images in the first and 6 images in the second set could further reduce the total number of images required to determine the fiber orientation with adequate accuracy.

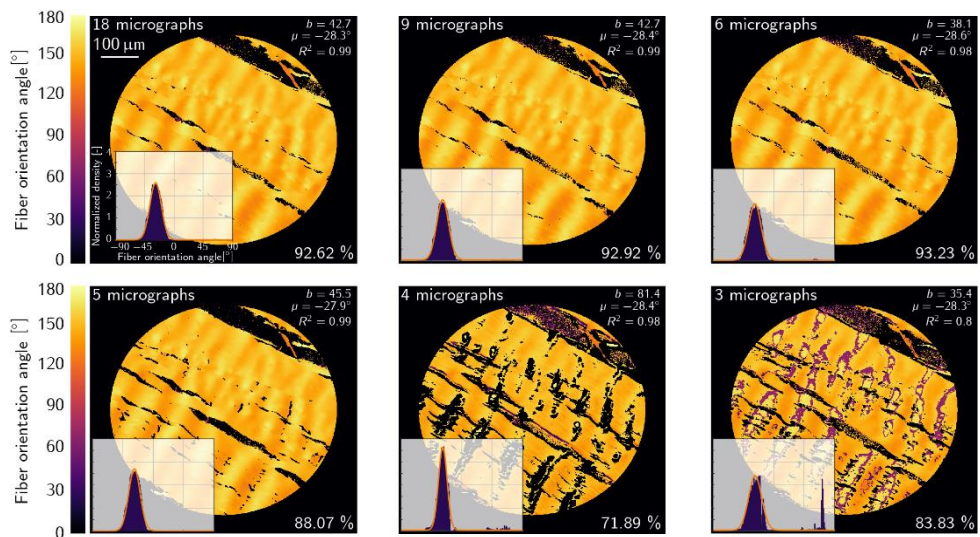

Fig. 5: Comparison of the resulting histograms and parameters of their von Mises approximations obtained with different numbers of input micrographs of the porcine Achilles tendon. The sample was intentionally oriented under the angle of approx. 152°= - 28°.

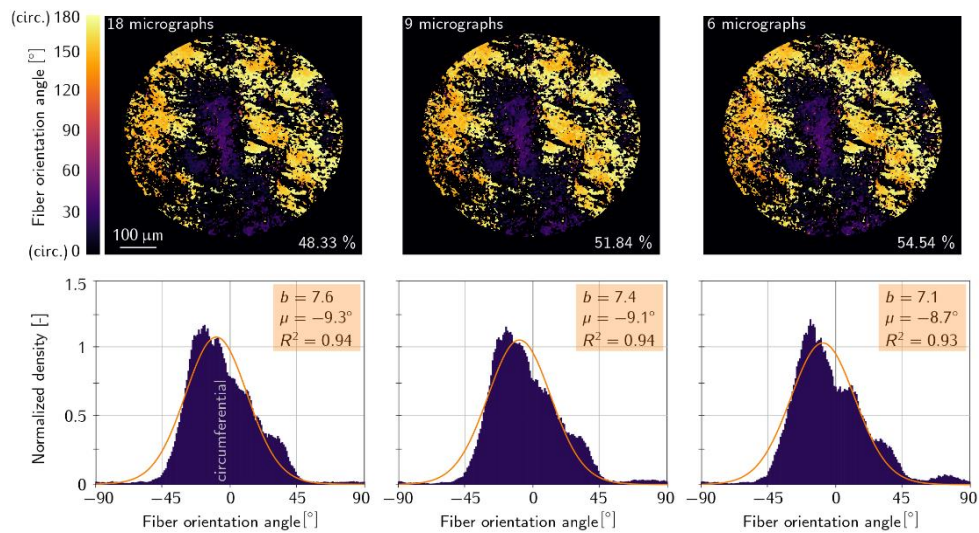

Fig. 6: Comparison of the resulting histograms of collagen directions in porcine aorta obtained with different numbers of input micrographs.

Finally, to check the suitability of the algorithm for tissues with a more complex collagen arrangement, we performed the same comparison for the media of porcine aortic wall. The evaluated sections are in the circumferential-axial plane with dispersion and waviness being much higher than in radial direction and thus more difficult for evaluation [11,29]. Here only some half of the pixels were evaluated, matching the percentage of collagen in this layer [30]. The histograms in Fig. 6 confirm that the mean angle of the distribution and its concentration parameter were evaluated correctly in all cases. We can conclude that 6 rotated micrographs are sufficient for evaluation of comparable tissues, ensuring no reduction in the number of evaluated pixels or incorrect evaluation of their directions.

## 4. Discussion

In this paper, we have proposed a new automated algorithm for evaluation of the orientation and dispersion of collagen fibers in soft tissues. It requires ca two minutes to record the needed 12 pictures of one micrograph with much lower requirements on the operator's experience and concentration, and then it takes a few seconds to evaluate the orientation in up to $5 \cdot 10^5$ pixels. The presented method overcomes two major limitations: time-consuming manual evaluation and the $\pi/2$-periodicity of light intensity at crossed polarizers. The used optical equipment - polarized light microscope with digital camera and 2D rotary stage - allowed us to evaluate reliably the orientation of collagen fibers in each pixel of the micrograph. For comparison, the study [24] used a manual method using PLM and evaluated a total of 5040 collagen orientations (30 different points at each of these slices).

To prove the correctness of the algorithm, it was verified using several manually evaluated pixels in micrographs of porcine Achilles tendon with mostly unidirectional and straight fibers. The subsequent validation against manual measurements with the same specimens demonstrated benefits and efficiency of the proposed technique. In the manual evaluation, we could not evaluate more than some 300 areas per microscopic image, while the algorithm determined the fiber orientation in each pixel, i.e. some $4.5 \cdot 10^5$ values in the same image. The much higher $R^2$ values of histogram approximations by von Mises distribution functions confirm a significantly better quality of structural parameters fitted to such histograms in comparison with the manual measurements. The robustness of the algorithm was highlighted also when the reduction of the needed number of rotated images was tested. The differences in structural parameters obtained on the basis of different sets of images were negligible (see Fig. 5) when at least 6 rotated images were used, thus establishing this number as optimum. This was the case not only with the highly uniformly oriented tendon tissue but also for the aortic wall with much more dispersed and curved collagen fibers. Although these histograms showed a visible asymmetry, their fitting with (symmetric) von Mises distribution function gave coefficient of determination $R^2 > 0.93$ in all cases and the resulting structural parameters were independent of the chosen number of images. Thus, the presented automated algorithm enables us to gain large data sets describing the in-plane distribution of collagen fibers in soft tissues and consequently also the structural parameters needed for the structure based constitutive description of collagenous tissue in computational modelling. The well-founded structural parameters significantly improve the quality of computational models of soft tissues, which in turn play an extremely important role in simulations of arteries under pathological conditions. As an example, these models facilitate the evaluation of rupture risk of aortic aneurysm [4,20] or fibrous cap of carotid atherosclerotic plaque [31] and may help in planning of surgeries.

In contrast to QPLM [17,18], CLSM [11,12] or MM [13], the proposed method does not require an expensive equipment; it is based on light microscopy and requires only the application of two polarizers and a standard rotary table. Some limitations remain, of course, mostly due to the basic principles of light microscopy. The thickness of specimens must be lower than 10 μm, thus their cutting with microtome may damage some structural components or tear fibers, especially if their out-of-plane dispersion or waviness are significant. The applied 5 μm thin slices are also very compliant and their position on the glass substrate may be distorted and thus vary locally. Also, we neglect the strains caused by unfolding of the rounded aortic specimens. Other errors could be introduced through misalignment of the polarizers or inaccurate positioning of the sample. A careful manipulation, however, may reduce these errors to a level insignificant in comparison with fiber dispersion and tissue variability. Despite these sources of errors, our procedure evaluated correctly the orientation of fibers, proving its robustness and effectivity. Naturally, errors related to extreme non-homogeneity of the tissue, such as in study [32], cannot be eliminated and must be treated by subdivision of the specimens.

The objective of the proposed method, similarly to most of the others, is to obtain histograms of fiber directions. For their following transformation into structural parameters of the constitutive models, we have shown here only approximation with unimodal von Mises distribution function. Naturally, for 3D distributions or for multimodal distributions (with more fiber families) more complex distribution functions has to be applied. However, it is not quite as easy to distinguish between fiber waviness and dispersion, and misinterpretations of wavy fibers as two fiber families may occur [27]. Thus completely different approaches should be adopted to obtain also parameters representing fiber waviness [13] that appear in some constitutive models (for instance [33]). These approaches may be based on histological investigation of the tissue under load, when the fibers are more or less straightened [31,32]; this issue is, however, out of scope of this paper. Transformation of histograms into structural parameters of constitutive models is addressed in another paper being just prepared.

In the context of our experiments, a higher resolution of the camera was not needed, since we do not study collagen fibers at the molecular level and the large view-field allowed us to quickly evaluate the examined area of the sample. Nevertheless, it is possible to easily enlarge the studied area using special objectives or a larger camera chip. Conversely, for applications in which high resolution is important, one can choose an objective with a high numerical aperture and magnification.

## 5. Conclusion

We demonstrated a new method enabling one an automatic, fast and accurate evaluation of orientation and dispersion of collagen fibers in soft tissues using polarized light microscopy. Our method is based on two sets of six rotated micrographs obtained with both perpendicular and inclined polarizers. The mutual inclination of both polarizers overcomes the $\pi/2$ ambiguity in the fiber orientation inherent to their orthogonal configuration because it is given by the $\pi/2$-periodic intensity function of the polarized light. Although dealing with in-plane distribution of collagen fibers, it enables also evaluation of their 3D distribution by combination of 3 mutually perpendicular sections. The prospect of this technique is to enable all the labs with standard equipment a fast evaluation of collagen orientation in the tissues. In human biomechanics, it will facilitate the evaluation of the structural constitutive parameters of blood vessel walls and other collagenous tissues which can be further used in computational modelling of their mechanical behavior under healthy and pathological conditions.

## Acknowledgement

This work was supported by Grant agency of the Czech Republic (https://gacr.cz/en/) projects No. 18-13663S and No. 20-01673S.

## Disclosures

The authors declare no conflicts of interest.

## Data Availability

Data underlying the results presented in this paper are not publicly available at this time but may be obtained from the authors upon reasonable request. Although the algorithm for fiber direction evaluation can be provided by the authors on request.

## References

1.  P. Fratzl, *Collagen: Structure and Mechanics, an Introduction* (Springer, 2008).
2.  S. G. Sassani, J. Kakisis, S. Tsangaris, and D. P. Sokolis, "Layer-dependent wall properties of abdominal aortic aneurysms: experimental study and material characterization," J. Mech. Behav. Biomed. Mater. **49**, 141–161 (2015).
3.  J. P. Vande Geest, M. S. Sacks, and D. A. Vorp, "The effects of aneurysm on the biaxial mechanical behavior of human abdominal aorta," J. Biomech. **39**, 1324–1334 (2006).
4.  T. C. Gasser, S. Gallinetti, X. Xing, C. Forsell, J. Swedenborg, and J. Roy, "Spatial orientation of collagen fibers in the abdominal aortic aneurysm's wall and its relation to wall mechanics," Acta Biomater. **8**, 3091–3103 (2012).
5.  G. A. Holzapfel, J. A. Niestrawska, R. W. Ogden, A. J. Reinisch, and A. J. Schriefl, "Modelling non-symmetric collagen fibre dispersion in arterial walls," (2015).
6.  A. C. Akyildiz, L. Speelman, and F. J. H. Gijsen, "Mechanical properties of human atherosclerotic intima tissue," J. Biomech. **47**, (2014).
7.  A. H. Hoffman, S. Louis, P. K. Woodard, S. Louis, K. L. Billiar, and L. Wang, "Stiffness Properties of Adventitia , Media , and Full Thickness Human Atherosclerotic Carotid Arteries in the Axial and Circumferential Directions," (2017).
8.  S. Polzer, T. C. Gasser, K. Novak, V. Man, M. Tichy, P. Skacel, and J. Bursa, "Structure-based constitutive model can accurately predict planar biaxial properties of aortic wall tissue," Acta Biomater. **14**, 133–145 (2015).
9.  A. Guinier, G. Fournet, and K. L. Yudowitch, "Small-angle scattering of X-rays," (1955).
10. R. T. Gaul, D. R. Nolan, and C. Lally, "Collagen fibre characterisation in arterial tissue under load using SALS," J. Mech. Behav. Biomed. Mater. **75**, 359–368 (2017).
11. J. T. C. Schrauwen, A. Vilanova, R. Rezakhaniha, N. Stergiopulos, F. N. van de Vosse, and P. H. M. Bovendeerd, "A method for the quantification of the pressure dependent 3D collagen configuration in the arterial adventitia," J. Struct. Biol. **180**, 335–342 (2012).
12. R. Rezakhaniha, A. Agianniotis, J. T. C. Schrauwen, A. Griffa, D. Sage, C. V. C. Bouten, F. N. Van De Vosse, M. Unser, and N. Stergiopulos, "Experimental investigation of collagen waviness and orientation in the arterial adventitia using confocal laser scanning microscopy," Biomech. Model. Mechanobiol. **11**, 461–

473 (2012).

13. A. Tsamis, J. A. Phillippi, R. G. Koch, S. Pasta, A. D'Amore, S. C. Watkins, W. R. Wagner, T. G. Gleason, and D. A. Vorp, "Fiber micro-architecture in the longitudinal-radial and circumferential-radial planes of ascending thoracic aortic aneurysm media," J. Biomech. **46**, 2787–2794 (2013).

14. E. Georgiou, T. Theodossiou, V. Hovhannisyan, K. Politopoulos, G. S. Rapti, and D. Yova, "Second and third optical harmonic generation in type I collagen, by nanosecond laser irradiation, over a broad spectral region," Opt. Commun. **176**, 253–260 (2000).

15. Y. Tanaka, E. Hase, S. Fukushima, Y. Ogura, T. Yamashita, T. Hirao, T. Araki, and T. Yasui, "Motion-artifact-robust, polarization-resolved second-harmonic-generation microscopy based on rapid polarization switching with electro-optic Pockells cell and its application to in vivo visualization of collagen fiber orientation in human facial skin," Biomed. Opt. Express **5**, 1099 (2014).

16. F. Massoumian, R. Juškaitis, M. A. A. Neil, and T. Wilson, "Quantitative polarized light microscopy," J. Microsc. **209**, 13–22 (2003).

17. J. C. M. Low, T. J. Ober, G. H. McKinley, and K. M. Stankovic, "Quantitative polarized light microscopy of human cochlear sections," Biomed. Opt. Express **6**, 599 (2015).

18. N. M. Kalwani, C. A. Ong, A. C. Lysaght, S. J. Haward, G. H. McKinley, and K. M. Stankovic, "Quantitative polarized light microscopy of unstained mammalian cochlear sections," J. Biomed. Opt. **18**, 026021 (2013).

19. C. E. Ayres, B. S. Jha, H. Meredith, J. R. Bowman, G. L. Bowlin, S. C. Henderson, and D. G. Simpson, "Measuring fiber alignment in electrospun scaffolds: a user's guide to the 2D fast Fourier transform approach," J. Biomater. Sci. Polym. Ed. **19**, 603–621 (2008).

20. S. Polzer, T. C. Gasser, C. Forsell, H. Druckmüllerova, M. Tichy, R. Staffa, R. Vlachovsky, and J. Bursa, "Automatic Identification and Validation of Planar Collagen Organization in the Aorta Wall with Application to Abdominal Aortic Aneurysm," Microsc. Microanal. 698–705 (2013).

21. J. P. McLean, Y. Gan, T. H. Lye, D. Qu, H. H. Lu, and C. P. Hendon, "High-speed collagen fiber modeling and orientation quantification for optical coherence tomography imaging," Opt. Express **27**, 14457 (2019).

22. H. M. Finlay, P. Whittaker, and P. B. Canham, "Collagen organization in the branching region of human brain arteries," Stroke **29**, 1595–1601 (1998).

23. A. J. Rowe, H. M. Finlay, and P. B. Canham, "Collagen biomechanics in cerebral arteries and bifurcations assessed by polarizing microscopy," J. Vasc. Res. **40**, 406–415 (2003).

24. P. Sáez, A. García, E. Peña, T. C. Gasser, and M. A. Martínez, "Microstructural quantification of collagen fiber orientations and its integration in constitutive modeling of the porcine carotid artery," Acta Biomater. **33**, 183–193 (2016).

25. L. C. U. Junqueira, G. Bignolas, and R. R. Brentani, "Picrosirius staining plus polarization microscopy, a specific method for collagen detection in tissue sections," Histochem. J. **11**, 447–455 (1979).

26. D. B. Chenault and R. A. Chipman, "Measurements of linear diattenuation and linear retardance spectra with a rotating sample spectropolarimeter," Appl. Opt. **32**, 3513 (1993).

27. K. Novak, S. Polzer, M. Tichy, and J. Bursa, "Automatic Evaluation of Collagen Fiber Directions from Polarized Light Microscopy Images," Microsc. Microanal.

**21**, 863–875 (2015).

28. T. C. Gasser, R. W. Ogden, and G. A. Holzapfel, "Hyperelastic modelling of arterial layers with distributed collagen fibre orientations," J. R. Soc. Interface **3**, 15–35 (2006).

29. S. Sugita and T. Matsumoto, "Multiphoton microscopy observations of 3D elastin and collagen fiber microstructure changes during pressurization in aortic media," Biomech. Model. Mechanobiol. **16**, 763–773 (2017).

30. M. K. O'Connell, S. Murthy, S. Phan, C. Xu, J. A. Buchanan, R. Spilker, R. L. Dalman, C. K. Zarins, W. Denk, and C. A. Taylor, "The three-dimensional micro- and nanostructure of the aortic medial lamellar unit measured using 3D confocal and electron microscopy imaging," Matrix Biol. **27**, 171–181 (2008).

31. W. Krasny, C. Morin, H. Magoariec, and S. Avril, "A comprehensive study of layer-specific morphological changes in the microstructure of carotid arteries under uniaxial load," Acta Biomater. **57**, 342–351 (2017).

32. S. V. Jett, L. T. Hudson, R. Baumwart, B. N. Bohnstedt, A. Mir, H. M. Burkhart, G. A. Holzapfel, Y. Wu, and C. H. Lee, "Integration of polarized spatial frequency domain imaging (pSFDI) with a biaxial mechanical testing system for quantification of load-dependent collagen architecture in soft collagenous tissues," Acta Biomater. **102**, 149–168 (2020).

33. G. Martufi and T. C. Gasser, "A constitutive model for vascular tissue that integrates fibril, fiber and continuum levels with application to the isotropic and passive properties of the infrarenal aorta," J. Biomech. **44**, 2544–2550 (2011).

**Supplemental document: Mathematical nature of polarized light microscopy**

The mathematical-physical basis of PLM is presented here in greater detail. The unpolarized light passes through the first polarizer P1 oriented along the x-axis, then through a sample characterized by a transmission matrix $\alpha$ (generally complex) and finally through the second polarizer P2 (see Fig. 1 b). The rotation of the sample coordinate system with respect to the global one is specified by angle $\theta$. The collagen fiber is implicitly assumed to coincide with the horizontal axis of the sample coordinate system (and with the orientation of the first polarizer), although generally, its orientation with respect to the sample coordinate system (in the main text denoted as *p*) is arbitrary and it is, in fact, the very parameter we strive to determine. Here, we set its value to zero to keep the notation simple and lucid, knowing it only leads to a shift in the angular intensity profile without any effect on its shape. The default orientation of the second polarizer is along the y-axis (the cross-polarized configuration of the microscope) and the deviation of P2 from this default orientation is denoted as angle $\delta$. The electric field vector $\boldsymbol{E}$ emerging from the setup can be calculated using the following matrix formalism:

$$\boldsymbol{E} = R^{-1}(\delta) P_2 R(\delta) R^{-1}(\theta) \alpha R(\theta) \boldsymbol{E}_0, \tag{S4}$$

where the individual matrices have the meaning as follows:

Incident wave (after passing the first polarizer):

$$\boldsymbol{E}_0 = \begin{pmatrix} 1 \\ 0 \end{pmatrix}, \tag{S5}$$

rotation matrix of the sample:

$$R(\theta) = \begin{pmatrix} \cos\theta & \sin\theta \\ -\sin\theta & \cos\theta \end{pmatrix}, \tag{S6}$$

rotation matrix of the P2:

$$R(\delta) = \begin{pmatrix} \cos\delta & \sin\delta \\ -\sin\delta & \cos\delta \end{pmatrix}, \tag{S7}$$

and for $\delta = 0°$ rotation matrix of the P2:

$$R(\delta = 0°) = \begin{pmatrix} 1 & 0 \\ 0 & 1 \end{pmatrix}. \tag{S8}$$

Transmission matrix:

$$\alpha = \begin{pmatrix} \alpha_{xx} & \alpha_{xy} \\ \alpha_{xy} & \alpha_{yy} \end{pmatrix}, \tag{S9}$$

$P_2$ matrix:

$$P_2 = \begin{pmatrix} 0 & 0 \\ 0 & 1 \end{pmatrix}. \tag{S10}$$

For the cross-polarized configuration ($\delta = 0$), the electric field attains the following simple form

$$\boldsymbol{E} = \begin{pmatrix} 0 \\ \frac{1}{2}(\alpha_{xx} - \alpha_{yy})\sin 2\theta + \alpha_{xy}\cos 2\theta \end{pmatrix}. \tag{S11}$$

In the experiment, the signal practically vanishes for values of $\theta$ equal to a multiple of $\frac{\pi}{2}$, therefore we will further neglect the off-diagonal terms of transmission matrix $\alpha$ (this means that the sample does not change the polarization state of the passing light when the collagen fibers are parallel or perpendicular to the polarization of the incident field). Consequently, the transmission matrix assumes a form coincident with a simple birefringent material.

The signal at the camera/detector is proportional to the intensity of the light emerging from the setup. Assuming the simplified expression for the transmission matrix, we obtain

$$\begin{aligned} I = |\boldsymbol{E}|^2 = &\frac{1}{8}|\alpha_{xx} + \alpha_{yy}|^2(1 - \cos 2\delta) \\ &+ \frac{1}{8}|\alpha_{xx} - \alpha_{yy}|^2(1 - \cos 4\theta \cos 2\delta - \sin 4\theta \sin 2\delta) \\ &+ \frac{1}{4}(|\alpha_{xx}|^2 - |\alpha_{yy}|^2)[\cos 2\theta\,(1 - \cos 2\delta) \\ &- \sin 2\theta \sin 2\delta]. \end{aligned} \tag{S12}$$

Although we use this expression in our calculations, it is insightful to take a limit for small values of $\delta$ (by expanding the trigonometric function into Taylor series and retaining only the lowest terms). Then the approximate expression for the light intensity reads

$$I(\theta) \approx \frac{1}{8}|\alpha_{xx} - \alpha_{yy}|^2[1 - \cos(4(\theta - p))]$$
$$-\frac{1}{2}\delta\left(|\alpha_{xx}|^2 - |\alpha_{yy}|^2\right)\sin(2(\theta - p)). \quad (S13)$$

The first term represents the basic profile obtained for the cross-polarized configuration ($\delta = 0$). Its periodicity is $\frac{\pi}{2}$ and one therefore cannot distinguish between the 45° and 135° orientations of the collagen fibers. Also note that it vanishes for $\alpha_{xx} = \alpha_{yy}$ which illustrates the necessity to use birefringent materials in these experiments. The second term has a periodicity of $\pi$ and for samples with $|\alpha_{xx}| \neq |\alpha_{yy}|$ it can account for the different spectra of colors observed for two perpendicular collagen fiber orientations.

By inspecting Eq. (S12) we can see that for one of the above orientations of collagen fibers (45°) the intensity of green color will increase while we should observe a decrease for the other one (135°), which nicely corresponds to the behavior observed in the experiment.

To verify this hypothesis, we used the full expression for the light intensity given by Eq. (S12) and compared our analytical model with measurements performed for different values of $\delta$ (see Fig. 2). Terms $|\alpha_{xx}|^2$ and $|\alpha_{yy}|^2$ can be measured directly using the configuration with aligned polarizers. The term $|\alpha_{xx} - \alpha_{yy}|^2$ can be obtained by fitting the measurements for $\delta = 0$. The last unknown quantity $|\alpha_{xx} + \alpha_{yy}|^2$ is then easily obtained from the equality

$$|\alpha_{xx} + \alpha_{yy}|^2 + |\alpha_{xx} - \alpha_{yy}|^2 = 2\left(|\alpha_{xx}|^2 + |\alpha_{yy}|^2\right). \quad (S14)$$